# A Cluster Controller for Transition Matrix Calculations

David Yevick

Department of Physics, University of Waterloo, Waterloo, ON, N2L 3G1
yevick@uwaterloo.ca

Yong Hwan Lee

Department of Physics, University of Waterloo, Waterloo, ON, N2L 3G1
yh24lee@uwaterloo.ca

## 1    Abstract

We demonstrate that a temperature schedule for single-spin flip transition matrix calculations can be simply and rapidly generated by monitoring the average size of the Wolff clusters at a set of discrete temperatures. Optimizing this schedule yields a potentially interesting quantity related to the fractal structure of Ising clusters. We also introduce a technique in which the transition matrix is constructed at a sequence of discrete temperatures at which Wolff cluster reversals are alternated with certain series of single-spin flip steps. The single spin-flip transitions are then employed to construct a single transition matrix.

## 2    Introduction

Biased sampling procedures such as the multicanonical [1] [2] [3] [4] [5] [6] [7] [8] [9] [10] [11] and Wang-Landau [12] [13] [14] techniques determine the statistics in low-probability regions of a small set of global variables $\vec{E}(\vec{\alpha})$ that depend on a generally large number of stochastically fluctuating local quantities $\vec{\alpha}$. These techniques are based on a Markov chain that is constructed by applying a small random change, $\delta\vec{\alpha}$ to the local parameter values yielding new quantities $\vec{\alpha}_{new} = \vec{\alpha}_{current} + \delta\vec{\alpha}$ $\vec{\alpha}_{new}$. Hence, specializing for simplicity to a single system variable, $E$ transitions from a histogram bin $i$ corresponding to an average energy $E_i$ to a (possibly identical) histogram bin $j$ with average energy $E_j$. The statistical bias is introduced by replacing $\alpha_{current}$ by $\alpha_{new}$ according to a acceptance rule that preferentially admits transitions toward physically relevant ranges of $\vec{E}(\vec{\alpha})$. At a subsequent point in the calculation the bias is appropriately removed.

Transition matrix procedures further enhance the accuracy of Markov chain biased sampling by recording both accepted and rejected transitions. That is, for every accepted or rejected transition from $E_i$ to $E_j$, the $i,j$:th element of an unbiased, unnormalized matrix $\tilde{T}$ is replaced by $\tilde{T}_{ij} + 1$. After the Markov chain steps are complete, each row in $\tilde{T}$ is normalized to unity. The resulting transition matrix elements $T_{ij}$ then coincide with the probability that the Markov chain evolves from a histogram bin $E_i$ to the bin $E_j$ in a single unbiased Markov step. [15] [16] [17] [18] [19] [20] [21] [22] [23] [24] [25] [26] The normalized unit eigenvector of $T$, which can be simply obtained either by repeatedly multiplying an initially random vector by $T$ [27] or by applying detailed balance to appropriate pairs of transition matrix elements, [16] [18] [22] [27] then coincides with the density of states (infinite temperature probability distribution) $p\left(\vec{E}(\vec{\alpha})\right)$ associated with $\vec{E}(\vec{\alpha})$. That the transition matrix method displays improved accuracy and scaling properties relative to its corresponding biased sampling technique has been exhaustively demonstrated in e.g. [28].

While the transition matrix is assembled from unbiased transitions and is therefore formally independent of the acceptance rule, its practical implementation is system-dependent. In the absence of phase transitions simple algorithms based on e.g. transition probabilities between microscopic states [16] [17], the ratio of transition matrix elements [17] [29] [30] [31], and the exclusion of transitions to bins that have been previously visited a larger number of times [23] [27] [32] yield results for $T$ that accurately reflect the underlying system behavior. However, near a phase transition large, connected regions of system variables with identical properties are present. As numerous small random displacements are required to evolve the system among such extended states, a sufficient number of Markov chain steps must be performed to sample correctly the cluster statistics. This can alternatively be viewed as the requirement to sample adequately the large number of configurations with differing values of the global variables accessible to the system near the critical point. [15]

The two-dimensional Ising model provides a simple and therefore relatively unambiguous context in which to compare the accuracy of different transition matrix procedures. In a single-spin flip procedure, $\delta\vec{\alpha}$ is associated with reversing a randomly selected spin. The standard procedure for performing such calculations biases the spin flips by employing the Metropolis acceptance rule [33] [34] to generate a Boltzmann energy distribution, although less accurate quasi-microcanonical methods exist that instead confine $\vec{E}(\vec{\alpha})$ to a narrow energy region. [35] [36] [37]. The matrix $\tilde{T}$ is then constructed while varying the inverse temperature (or the quasi-microcanonical energy) between a near zero and a suitably large value such that $\beta \equiv 1/T = f(m)$ in the m:th computation step, where $f(m)$ is a generally monotonic function, often termed the inverse temperature schedule, that interpolates between the initial and final inverse temperatures and varies slowly near the inverse critical temperature. Numerous proposals have been advanced for optimizing $f((m)$ based on, for example, the correlation time, [28] the distribution of realizations over magnetizations [36] and the entropy of either the full phase (configuration) space or the phase space in the energy-magnetization $(E - M)$ diagram. [37] Additionally, the numerical accuracy can be enhanced by accumulating transitions from multiple independent Markov chains or through a renormalization procedure that infers an approximate density of states from the corresponding result for a smaller system. [28] [36] [38]

In a previous paper, the authors advanced a new approach to transition matrix calculations that combined the Wolff algorithm with the Markov chain single spin flip procedure.[27] While this analysis was limited to the Ising model, the Wolff procedure can be replaced by machine learning approaches for more complex physical systems. [39] [40] To implement a representative version of the integrated method, while $\tilde{T}_{ij}$ is assembled in standard fashion by reversing the spin of randomly selected spins, global cluster flips generated by the Wolff method [41] (or any analogous cluster inversion procedure) are interspersed with certain of these transitions. Successive Wolff realizations sample nearly uncorrelated system configurations in contrast to a Markov chain based on single spin-flips, which instead diffuses slowly through phase space. Therefore, following a cluster flip with single spin flips insures that the states employed to construct the transition matrix even after a relatively small number of computation steps represent those of the equilibrium ensemble. However, as the Wolff steps are computationally intensive close to and especially below the critical temperature, $T_c$ while away from $T_c$ the extent of the accessible phase space and hence the computation time of the single spin flip method is greatly reduced, [15] the Wolff procedure was only sparingly employed in a region near the critical temperature.

In this paper, we extend our previous analysis by noting that the Wolff procedure for a fixed temperature yields after a small number of steps a reliable estimate of the mean cluster size (the method can also be applied as the temperature is continuously varied but the cluster size will then typically be averaged over a narrow range of temperatures). Since the cluster shape, especially near the transition temperature, has been shown to have a fractal dimension, the rate at which a single step procedure will effectively fully modify an entire cluster will be a non-integer power of the number of spins in the cluster. This power, which could constitute a possibly novel fractal quantity, will be estimated by referring to a previous determination of the optimal temperature schedule based on monitoring the coverage of the accessible $E - M$ phase space. [37] Once the relationship between the cluster size and the temperature schedule is established, a single spin-flip transition matrix calculation based on the cluster schedule is compared with a similar calculation that, however, judiciously intersperses Wolff cluster reversals with single spin-flips.

## 3 Numerical Methods

The numerical procedures in this paper incorporate both the Wolff and single spin-flip algorithms. The Wolff method starts from a random spin in a lattice system at temperature $T$ and determines the largest connected region surrounding the spin containing spins with the same orientation. Subsequently all spins in this cluster are reversed. Repeating this numerous times generates a set of states that possess a Boltzmann distribution of energies at $T$. While very rapid for high temperatures and hence small cluster sizes, the method is highly inefficient at low temperatures for which a single cluster extends throughout nearly the entire computational window. The single spin-flip procedure on the other hand, reverses a single, randomly selected spin in an "old" system realization, generating a potential transition from $E_{\text{old}}$ to $E_{\text{new}}$. In the transition matrix method the element of $\tilde{T}$ associated with a transition from the bin containing $E_{\text{old}}$ to the bin $E_{\text{new}}$ is then incremented by unity. Subsequently this proposal is accepted with a probability $p_{\text{acceptance}} = \min(e^{\beta(E_{\text{old}} - E_{\text{new}})}, 1)$ .

The large phase space displacements generated by the Wolff algorithm are achieved at the cost of long computation times especially for temperatures near or below the critical temperatures. In a previous publication, the latter drawback was addressed by interspersing Wolff and single spin-flip steps in a temperature interval near the critical temperature while accumulating in $\tilde{T}$ only the spin flip transitions [27]. Here, however, several of the following more accurate procedures will instead be examined or implemented.

1. The Wolff procedure can be implemented at discrete temperatures $T_n = T_0 + n\Delta T$, $n = 1,2,\ldots N$ and the resulting transitions for each calculation stored in individual matrices $\tilde{T}^{(n)}$. The unit eigenvector of the associated matrix $T^{(n)}$ then yields the partial density of states within a region of $\beta$ centered around $\beta_n$. These partial results can then be combined to form the full density of states as in [37] [42] However, this methods is limited by the inefficiency of the Wolff algorithm at low temperatures.

2. Following the strategy of [27], each Wolff cluster reversal in the above procedure can be followed by $N_s$ single spin flip states and only the transitions generated by the single spin flips stored in a single matrix $\tilde{T}$. As illustrated below, besides eliminating the patching step, this affords a broader coverage of $E$ for a rather insignificant increase in computation time.

3. A small number, $N_{Wolff}$ of Wolff steps can be performed at each $\beta_n$. At each step, the Wolff algorithm identifies a single cluster and can therefore return its size (e.g number of spins). An average is then performed over all $N_{Wolff}$ cluster sizes. If the number of spins in the lattice and the average number of spins in a Wolff cluster are denoted $N_{spins,total}$ and $N_{average,cluster}$, respectively, identifying the relevant cluster size below the transition temperature with that of the non-dominant spins, the expression

$$N_{spin\ steps}^{(j)} = N_{ref} * \left|\frac{N_{spins,total}}{2} - \left|N_{average,cluster}^{(j)} - \frac{N_{spins,total}}{2}\right|\right|^{\alpha} \quad (1)$$

(which is proportional to the temperature schedule) is employed to determine the required number of single spin flip steps at each $\beta_n$. Appropriate rounding functions are applied to insure that the number of spin steps is non-zero and that Eq.(1) evaluates to an integer. The value of $\alpha$, as explained below, is further chosen such that the temperature schedule nearly coincides with that of [37]. The single-spin flip transitions for all $\beta_n$ are collected in a single transition matrix

4. Through interpolation, the temperature schedule associated with Eq.(1) can also be employed in a calculation with a continuously varying temperature as in [27].

## 4   Results

The accuracy and efficiency of the above algorithms will now be established through a benchmark determination of the $32 \times 32$ spin Ising model specific heat for zero external magnetic field, periodic boundary conditions, and a unit amplitude ferromagnetic interaction. To illustrate the relative performance of the Wolff and single spin flip algorithms, Figs.(0) and (1) display the exact (thin solid line) and estimated logarithm of the number of states as a function of energy in units of $J/k_b$ for normalized temperatures of $T = 18$ and $T = 2.4$, respectively where $J$ represents the spin-spin interaction energy. Here the + markers are associated with method (1) of the previous section in which $1024 \times 10^3$ Wolff cluster reversals are applied to an initially thermalized system. The thick solid line is instead obtained with method (2) by following each of these Wolff cluster reversal by 1000 single spin flips and populating $\tilde{T}$ with only the single spin flip transitions. The computation time is nearly identical in the two calculations. Clearly the integrated Wolff/single spin flip procedure (method 2 above) yields the partial density of states over a wider energy range compared to a method based solely on Wolff steps, although the advantage of the additional single spin flips is reduced near a phase transition. More significantly, however, integrating single spin flips with the Wolff method enables the transitions from different temperature regions to be collected in a single transition matrix, considerably reducing the programming effort required to construct the full density of states.

Next, to verify that the Wolff procedure yields a stable result for the average number of spins in a cluster, Fig.(2) displays the number of spins in a cluster averaged over differing numbers of Wolff steps for $T = 2.26$, where $T_c = 2.269$ is the thermodynamic critical temperature for an infinite system size. [43] [44] [45] Evidently, averaging the cluster sizes generated by $\approx 2000$ Wolff steps yields an accurate estimate equal to approximately half the spins in the system. The average cluster size can therefore be generated very rapidly in practical calculations after which Eq.(1) yields the preferred number of single spin flip steps to be executed at each temperature in a combined Wolff/single spin flip calculation. With $\alpha = 0.8$ the schedule, Fig.(3), obtained after 10000 Wolff steps at each of a grid of temperatures spaced by 0.1 in normalized units of $J/k_B$ and by 0.01 within the single 0.1 temperature interval containing $T_c$ nearly coincides with that of Fig.(13) of [37]. To be consistent with the normalized units of [37], the horizontal axis in this figure displays the quantity $4/T$ while the vertical dashed line denotes $4/T_c$. Note that since [37] the optimal schedule is generated from the number of single spin flip steps at each temperature required for the Markov chain to sample evenly the entire accessible configuration space in the $E - M$ plane, $\alpha = 0.8$ would appear to imply that the number of steps required to reconfigure a cluster is governed by a possibly novel fractal dimension. That is, if the clusters were circular, their radius would vary as the square root of the number of spins within a cluster. In this case, if the change in cluster size were proportional to the number of single spin-flip steps, $\alpha$ in Eq. (1) would optimally equal 0.5. On the other hand, if the reduction in radius of the initial cluster is governed by a diffusion process the reconfigured spins would spread inward from the edge of the cluster at a rate given by square root of the number of single spin flips yielding $\alpha = 1$. Because of the fractal form of the cluster interface, however, the clusters are eliminated at a rate that interpolates between these two values.

Having established a near-optimal temperature schedule, specific heat benchmark calculations can be performed. The first result, Fig(4) displays (dashed line) the average over 10 specific heat curves obtained by accumulating in a single matrix the spin flip transitions at each of the discrete temperatures

$T^{(n)} = 18.0, 10.0, 6.5, 5.0, 3.8, 3.1, 2.85, 2.6, 2.4, 2.26, 1.99, 1.82, 1.45, 1.2$ together. Since the result is almost indistinguishable from the solid line (exact result) in the figure, the insert displays the similarly averaged percentage relative error defined as $|(C(T) - C_{exact}(T))/C_{exact}(T)|$ The number of single spin flips at each temperature was obtained from Fig(3) together with $N_{ref} = 1.e7$. The computation time for each curve was 15.3 minutes on an Intel i7 processor. For comparison, this computation is repeated in Fig.(5) except that 10,000 Wolff steps are interspersed with the single spin flip steps for each of the temperatures 2.6, 2.4, 2.26 and 1.99. While the total number of single spin flips is identical in the two figures, the additional Wolff steps increase the total computation time by 82%. Comparing Fig.(4) and Fig.(5) suggests, as previously noted in [27], that combining the Wolff and single spin flip methods can significantly enhance the precision of practical calculations.

## 5   Discussion and Conclusions

While the main result of this paper, namely that the Wolff algorithm can be employed to simply and efficiently generate a temperature schedule for single spin-flip calculations has obvious practical implications, several theoretical issues of potential significance remain. First, the Wolff algorithm rapidly evaluates the average size of a spin cluster at a given temperature while in a previous paper [37] an optimal temperature schedule for single-spin flip calculations was derived based on an analysis of the number of single spin-flip steps required to sample evenly the accessible states in the energy-magnetization plane. Combining these two approaches led to an estimate of the fractal power dependence of the effective cluster radius on the number of states that it contains. While extensive analysis would be required to optimize the temperature schedule and to describe precisely the process of cluster formation and evolution, this could potentially lead to the identification of a new fractal property.

Another aspect of the cluster controller is that, as noted in a previous paper, in more complex systems a direct analog either to the Wolff method or to analogous cluster inversion procedures [46] [47] [48] may be absent. Although machine learning techniques that map the physical system near the critical temperature onto a more easily manipulated model could then be employed [39] [40] in place of the Wolff algorithm, these do not necessarily directly predict the cluster size. However, a useful temperature schedule could still be presumably obtained by examining, for example, the width of the central peak of the fast Fourier transform of the spin distribution. In this manner, the cluster controller can be viewed as isolating the determination of the optimal temperature schedule from the details of the physical system.

## Acknowledgments

The Natural Sciences and Engineering Research Council of Canada (NSERC) is acknowledged for financial support.

## 6 Figures

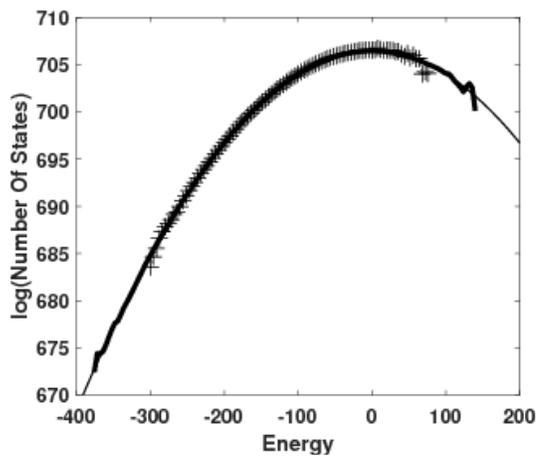

Figure 1: The exact (thin solid line) logarithm of the number of states as a function of energy in units of $J/k_b$ for temperatures of $T = 18$ and $T = 2.4$ in units of $k_b$ The + markers are the result of a transition matrix calculations populated by Wolff transitions (method 1) while The thick solid line is generated by following each Wolff cluster reversal with 1000 single spin flips and storing only the single spin flip transitions.

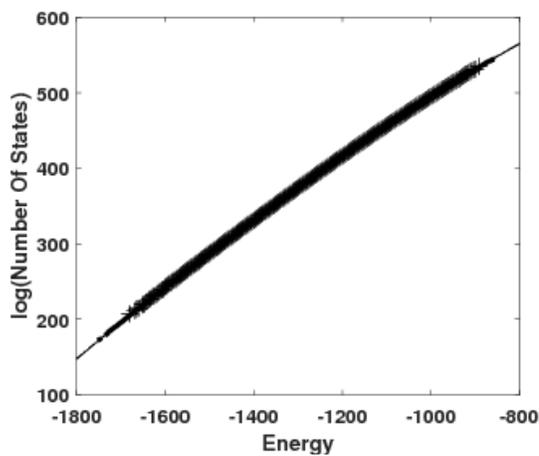

Figure 2: As in Fig(0) but for $T = 2.4$.

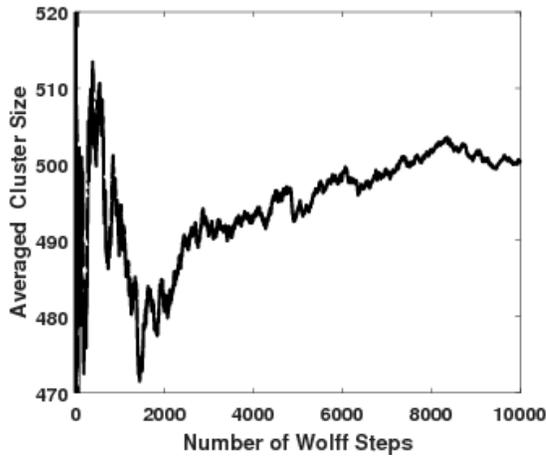

Figure 3: The averaged number of spins in a Wolff cluster at $T = 2.4$ as a function of the number of Wolff steps in a representative calculation.

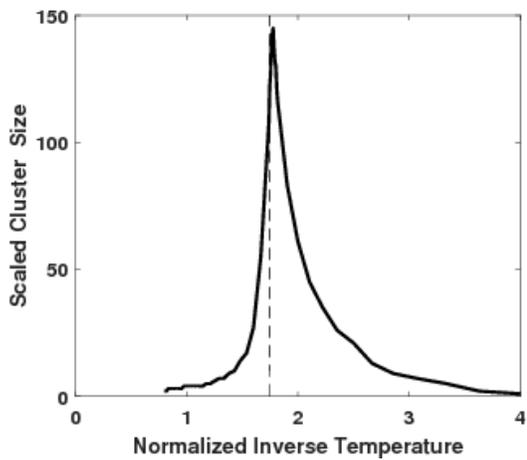

Figure 4: The temperature schedule generated from the Wolff cluster size combined with an analysis of the state diffusion in the E-M plane. The dashed vertical line denotes the inverse thermodynamic critical temperature while the horizontal axis represents $4/T$ with $T$ in units of $J/k_B$ where $J$ is the spin-spin interaction strength.

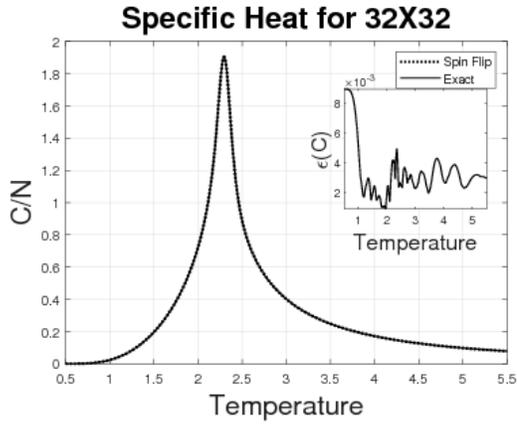

Figure 5: The average over 10 specific heat per spin curves versus temperature, obtained with the cluster controller for a transition matrix populated by exclusively single spin flips at each of the discrete temperatures cited in the text. The inset shows the averaged relative error.

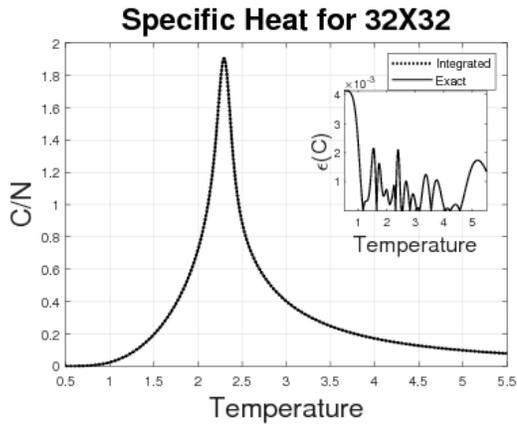

Figure 6: As in Fig.(4) but with 40,000 Wolff cluster reversals interspersed with the single spin flip Markov chain at the four temperatures closest to the critical temperature